# MAEC: A Movement-Assisted Energy Conserving Method in Event Driven Wireless Sensor Networks

Ming Zhao, Zhigang Chen, Xiaoheng Deng, Lianming Zhang, Anfeng Liu, and Guosheng Huang
College of Information Science and Engineering, Central South University
Changsha, Hunan, 410083, China
Email: meanzhao@gmail.com, czg@csu.edu.cn

**Abstract:** Energy is one of the most important resources in wireless sensor networks. Recently, the mobility of base station has been exploited to preserve the energy. But in event driven networks, the mobility issue is quite different from the continuous monitoring one because only a small portion of sensor node has data to send at one time. The number of sensor node that forward traffic should be minimized to prolong the network lifetime. In this paper, we propose a movement-assisted energy conserving method which tries to reduce the amount of forwarding sensor node by directing the base station to move close to the hotspots. This method achieves good performance especially when applied to a network with a set of cooperative mobile base station. Extensive simulation has been done to verify the effectiveness of the propose schema.

## Ⅰ. INTRODUCTION

The recent advances in wireless communications and electronics are paving the way for the deployment of low-cost, low-power networks of untethered and unattended sensors. Energy efficiency is one of the most important issues in the design of wireless sensor networks. In most cases, sensor nodes are small and cheap, and have very limited battery capacity. Moreover, they are usually deployed in large quantity and in some cases the physical environment surrounding them could be inhospitable. Thus, replenishing the sensor's battery is not cost-effective, and may even be impossible in practice.

Corresponding to the importance of the problem, there has been vast research addressing different aspect of the power control problem in wireless sensor networks. These include, among others, energy efficient MAC and routing protocols (e.g., [1], [2], [3], [4] and [5]), low-power security protocols (e.g., [6], and [7]), clustering (e.g., [8], [9], and [10]), data aggregation (e.g., [11], [12], and [13]) and scheduling (e.g., [14]).

In ref. [15], Madden et al differentiate between event driven and continuous monitoring sensor networks. The differentiation is mainly dictated by the application. In an event driven sensor network, the sensor nodes do not send data (and are most likely asleep) until a certain event occurs. For example, consider an application of monitoring toxicity levels in an area in which hazardous materials are used and hazardous waste is produced. Until toxicity is detected, no data needs to be sent.

In event driven sensor networks, the energy consumption of sensor nodes situating outside the "hotspots" (the area where a certain event occurs) is attributed to forwarding data. Intuitively speaking, the farther the base station lays away from the hotspots, the more the sensor node that has to forward traffic. In another word, the number of forwarding sensor node will be greatly reduced if the base station can move close to the hotspots.

Although the base station is usually assumed to be static, it can become mobile thanks to the advance made in the field of robotics. Recently, the mobility of base station has also been exploited to extend the network lifetime (e.g., [16], [17], and [18]). Unfortunately, all these work achieve their optimization under continuous monitoring networks, and their results are difficult to be generalized to event driven networks. To the best of our knowledge, there is little research about this issue.

The rest of the paper is organized as follows. In section Ⅱ, we discuss related work to place our contributions in context. Section Ⅲ analyzes sensor networks with a static base station and with a mobile base station. Next, in section Ⅳ and Ⅴ, we propose the movement-assisted method. Simulation is presented in section Ⅵ. Section Ⅶ provide the concluding comment and future work.

## Ⅱ. RELATED WORK

The existing work concerned with energy conservation issues is so vast that it would deserve several comprehensive surveys. Hence, we only discuss a few related topics here. Energy efficient routing protocols (e.g., [1], [2], [3], and [4]) try to maximize the network lifetime by balancing the energy consumption among all nodes. Energy efficient MAC protocols (e.g., [5]) aim at reducing energy consumption and support self-configuration. Security in wireless sensor networks is difficult because of the limited processing power, storage, bandwidth, and energy. Low-power security protocols (e.g., [6], and [7]) address this issue. Clustering (e.g., [8], [9],

This work was supported in part by the National Research Foundation for the Doctoral Program of Higher Education of China (200405333036), the Foundation for University Key Teacher by the Ministry of Education project ([2000]143), and the Natural Science Foundation of Hunan Province (03JJY4054).

and [10]) and data aggregation (e.g., [11], [12], and [13]) are usually closely correlated [8]. Applications requiring efficient data aggregation (e.g., computing the maximum detected radiation around an object) are natural candidates for clustering. In-network aggregation (fusion) of data packet enroute to the base station is performed by selecting a set of cluster heads among the nodes in the network. The key idea is to combine data from different sensor nodes to eliminate redundant transmission. Because the best way to conserve energy is to turn the sensor nodes off, cross-layer scheduling (e.g., [14]) aims at forming on-off schedules to enable the time-synchronized sensor nodes to be awake only when necessary. While all these results appear to be somewhat different to the work in this paper, they are all potentially complementary to our idea.

With the development of mechanical technology, there have been exist several research efforts on exploiting the mobility of base station to elongate the network lifetime (e.g., [16], [17], and [18]). Ref. [16], [17] adopt the mobile relay approach, which will lead to significant delays of data transmission. Ref. [18] analytically quantifies the benefit in terms of network lifetime due to mobility, and proposes that the best mobility strategy consists in following the periphery of the network. But this method depends heavily on the shape of the network region and difficult to be applied to event driven network.

III. TO MOVE OR NOT TO MOVE

To verify our intuition, we compare the energy consumption of network with static and mobile base station respectively in this section. In one case, the base station stay still at all times. In another case, we require the base station to move close to the hotspots. Our test is implemented with the well-know simulator NS-2 (version 2.1b9a).

We assume a relatively dense and strongly connected network with each sensor knows its position (through GPS). There are 200 sensor nodes randomly deployed in a 750m×750m field. The base station and the hotspots are indicated as Fig. 1 (red × denotes base station and red ellipse stands for hotspots).

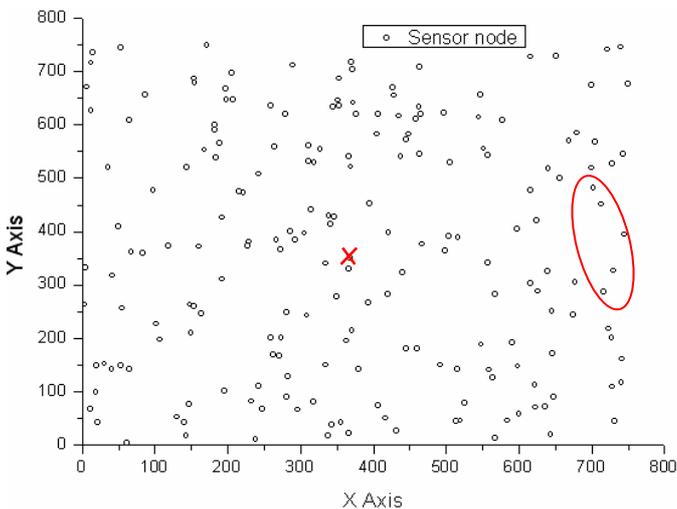

Fig. 1. Topology of our test event driven wireless sensor network

*A. Base Station Stay Still*

If we fix the base station, then after 300 packets been send (by each node in hotspots), the remaining energy is plotted in Fig. 2.

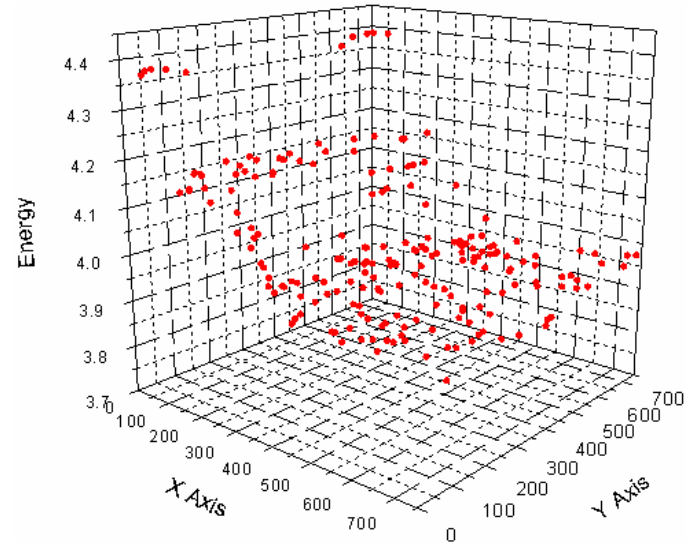

Fig. 2. Sensor node residual energy distribution with base station stay still (initial energy set 4.455J)

Fig. 2 shows the energy consumption of sensor nodes which lie between hotspots and base station (whose x-coordinate belongs to [350, 750]) is very high comparing to that of the other side thanks to the forwarding, collision and retransmission.

Beside, it is easy to see that all sensor nodes have consumed a certain part of energy, which is due to the route discovery and maintenance process.

*B. Base Station Move Close to Hotspots*

In this case, we manually move base station to the center of hotspots after every sensor node have sent a packet (to simulate the scenario that base station has detect the hotspots). As in the previous case, after 300 packets been send (by each node in hotspots), the remaining energy is displayed in Fig. 3.

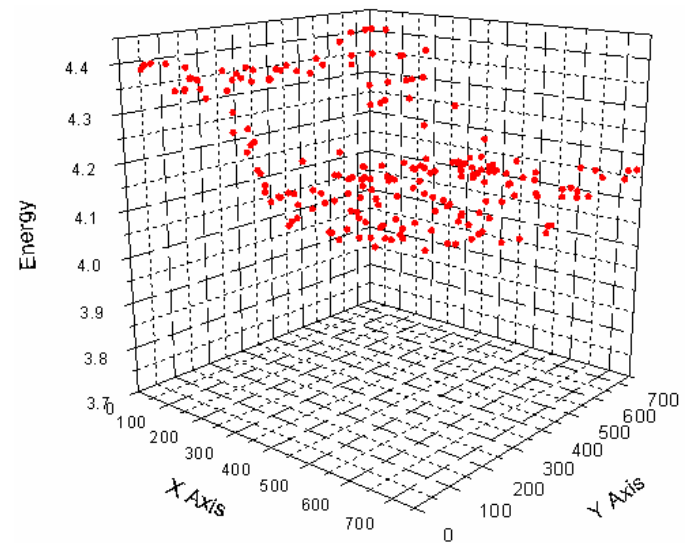

Fig. 3. Sensor node residual energy distribution with base station move close to hotspots (initial energy set 4.455J)

Obviously, energy consumption is much lower than the previous case, although the consumption of sensor nodes whose x-coordinate belongs to [350, 750] is still higher than that of the others because of collision and retransmission. In real world networks, this portion of consumption can be further reduced with appropriate MAC, routing or scheduling measures.

*C. Comparison*

By comparing Fig.2 and Fig.3, we can easily find out the obvious difference of two cases (base station stay still and move close to hotspots). The second case consumed much less energy than the first one, thanks to the movement of base station which reduce the transmission hops dramatically. Fig.4 and Fig.5 reveal the difference much more explicitly.

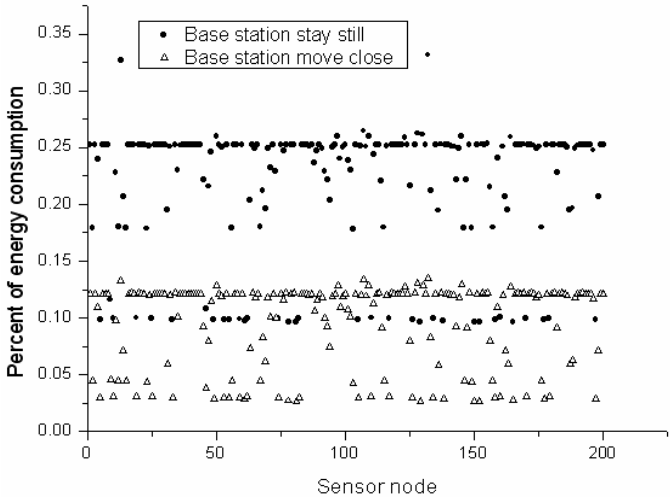

Fig. 4. Comparison of energy consumption percent of each sensor node in two cases

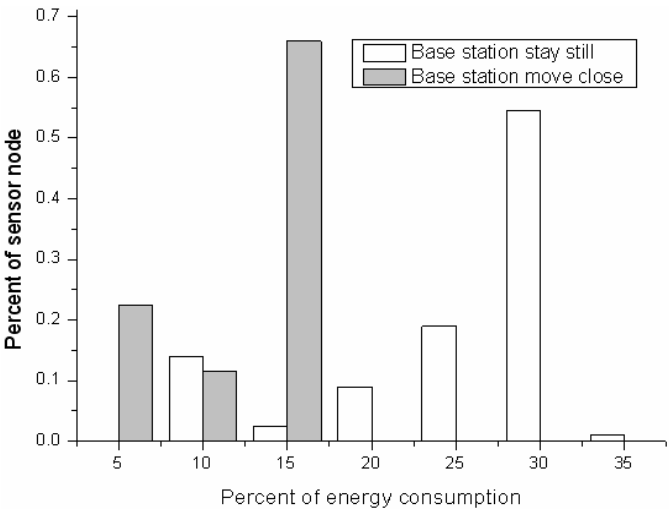

Fig. 5. Comparison of sensor node distribution with different energy consumption percent (two scenarios)

The above two figures tell that energy consumption with base station moving close is nearly half of that with static base station, which demonstrate that our intuition is indeed a promising strategy to prolong the lifetime of event driven networks.

*D. Another Issue*

Intuitively speaking, because of the specialty of event driven network, the benefit of even load getting from mobility [18] may not hold any more. To the contrary, it will lead to more energy consumption because much more sensor nodes has to take the responsibility of forwarding data. Simulation has also been done to prove our intuition, the result is displayed in Fig. 6 (just the same as A & B, 300 packets been send by each node in hotspots).

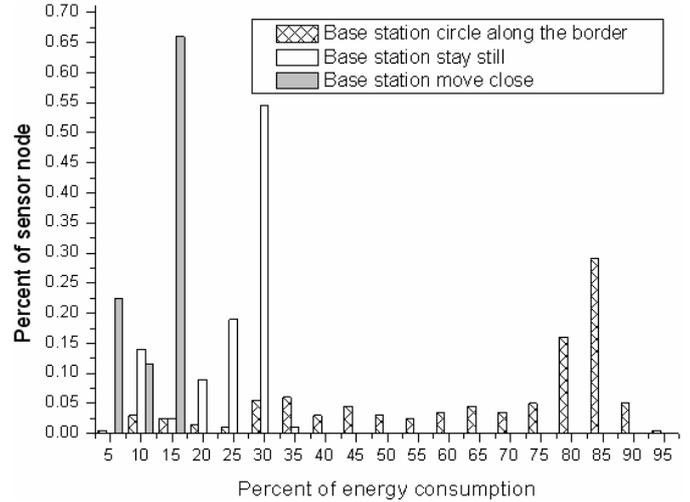

Fig. 6. Comparison of sensor node distribution with different energy consumption percent (three scenarios)

Above figure argues that the way base station move should be well designed, which is the main object of next section.

## IV. MAEC

In the previous section, we have shown that mobility helps to conserve energy and prolong the event driven sensor network lifetime. Now the question is how to move? This section introduces the MAEC (Movement-Assisted Energy Conserving) method.

*A. General Idea*

The results of previous section motivate us to design a mechanism to control the movement of base station. We view each sensor node generate a virtual force to attract the base station. Because of several virtual forces is working at the same time, hence the base station is driven by the composition of forces.

We suppose that the location and size of hotspots do not change fast, i.e., the highly dynamic environment is out of the scope of this paper, which is left to our future work.

We divide the whole space into eight sub-spaces (①, ②, …, ⑧), For example, in Fig. 7, there are two source sensor nodes (sensor node which initiate a data transmission) in sub-space ③, four in ④ and one in ⑤. And as in ref. [19], a sensor node which communication radius $r$ is said to be physical $i$ hops away from the base station if the Euclidean distance between the sensor node and the base station belongs to $((i-1)r, ir]$.

For example, in Fig. 7, there are three sensor nodes physical 2 hops away and four physical 3 hops away.

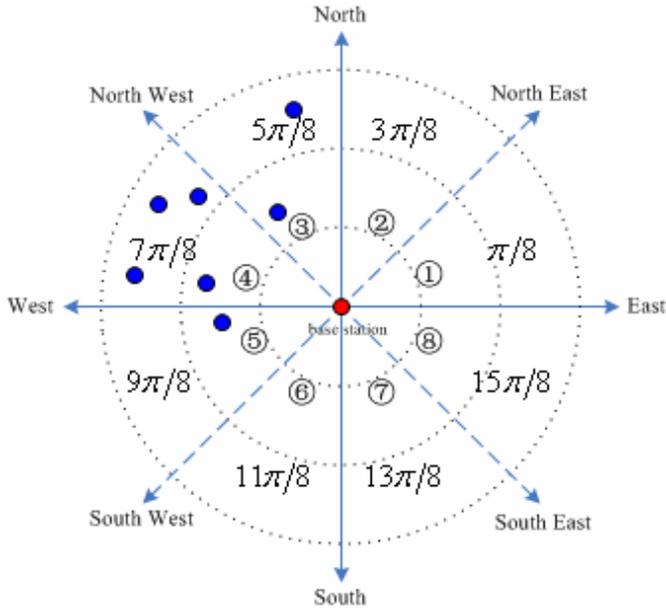

Fig.7. The space division and physical hop example

*B. Direction to Move*

The direction of base station is calculated by

$$Direction = \frac{1}{N}\sum_{i=1}^{4} C_i, \quad \quad (1)$$

where $c_i = \begin{cases} (h_i n_i - h_{i+4} n_{i+4})(2i-1)\pi/8 & h_i n_i - h_{i+4} n_{i+4} > 0 \\ (h_{i+4} n_{i+4} - h_i n_i)[2(i+4)-1]\pi/8 & h_{i+4} n_{i+4} - h_i n_i > 0 \end{cases}$,

$n_i$ is the number of source sensor node lies in sub-space $i$, $h_i$ is the average physical hops of source sensor node in sub-space $i$, and $N = h_{avg} \sum_{i=1}^{4} |n_i - n_{i+4}|$, where $h_{avg}$ is the average physical hops of all source sensor nodes.

For example, the direction of base station in Fig.7 is

$$\frac{1}{7 \times 18/7} \sum \begin{bmatrix} 2\times 1 \times 9\pi/8 \\ +2.5 \times 2 \times 5\pi/8 \\ +2.75 \times 4 \times 7\pi/8 \end{bmatrix} = 5\pi/6$$

*C. Distance to Move*

The distance of base station move is decided by the following equation:

$$\frac{1}{N}\sum \left[ (i-1) \times n_i \times r \right], \quad \quad (2)$$

where $r$ is the communication radius, $n_i$ is the number of source sensor node with physical hops $i$, and $N=\sum n_i$.

For example, in Fig. 7, there are three source sensor nodes with physical hops 2, four with physical hops 4. So the distance of base station shall move is

$$\frac{1}{7}\sum \begin{bmatrix} (2-1)\times 3 \times r \\ +(3-1)\times 4 \times r \end{bmatrix} = 11r/7$$

*D. The MAEC*

A formal description of MAEC is shown in Fig. 8. The MAEC method runs round by round. Each round consists of three phases: *waiting*, *discovery* and *moving*. In the *waiting* phase, the base station set the direction and distance to be zero and wait for incoming packets.

---

*Notations:*
*source⟨s_loc⟩*: the location of a source sensor node
*list*: list of *source⟨s_loc⟩*

**1** *Upon entering Waiting phase:*
   **1.1** set *direction* and *distance* to be zero
   **1.2** enter *Discovery phase* upon receiving data
**2** *Upon entering Discovery phase:*
   **2.1** set *timer* to be *discovery_interval*
   **2.2** upon receiving data from a new *source⟨s_loc⟩*, add it to *list*
   **2.3** upon timeout compute the *direction* and *distance* of next move base on *list* using equation (1) and (2)
      **2.3.1** enter *Waiting phase* if no data received, or else enter *Moving phase*
**3** *Upon entering Moving phase:*
   **3.1** set *timer* to be *moving_interval*, enter *Discovery phase* upon timeout
   **3.2** move as *direction* and *distance* required

Fig.8. The MAEC algorithm

## V. CoMAEC

When there are several hotspots located around the base station, then the MAEC will possibly actually decide to stay static (the direction and distance being too small). That is to say, the difference between MAEC and static will decease.

Fortunately, there are cases when there are a set of base station available in network. In these networks, every sensor's messages must be routed to some base station, where the data can be processed [20]. The initial layout of base station set can be randomly or decided by method in [20].

The MEAC in this paper can be easily extended to be applicable in this scenario, which we call the CoMAEC (cooperative MAEC) method. The key problem of interest is that of how to effectively control the movement of each base station. It is straight that energy consumption achieve optimal if there is one base station lie inside each hotspots. Unfortunately, in general, there are more hotspots than base stations.

To decide the movement of each base station, one base station is selected as the head (using round robin or other election methods) and is responsible for assign each base station a group of source sensor nodes.

The CoMAEC method runs round by round as described in Fig. 9. Each round consists of four phases: *waiting*, *discovery*, *clustering* and *moving*.

*Notations:*
source⟨s_loc⟩: the location of a source sensor node
*list*: list of source⟨s_loc⟩

**1** *Upon entering Waiting phase:*
   **1.1** set *direction* and *distance* to be zero
   **1.2** enter *Discovery phase* upon receiving data
**2** *Upon entering Discovery phase:*
   **2.1** set *timer* to be *discovery_interval*
   **2.2** for each base station, upon receiving data from a new source⟨s_loc⟩, add it to *list*
   **2.3** upon timeout enter *Waiting phase* if no data received, or else enter *Clustering phase*
**3** *Upon entering Clustering phase:*
   **3.1** each base station send its own *list* to the base station head
   **3.2** base station head divide the source sensor node into several groups depending on the number of base station
   **3.3** base station head assign each base station a group of source sensor nodes depending on their distance to the group
   **3.4** for each base station calculate the direction and distance of next move base on its new *list* using equation (1) and (2)
   **3.5** enter *Moving* phase
**4** *Upon entering Moving phase:*
   **4.1** set *timer* to be *moving_interval*, enter *Discovery phase* upon timeout
   **4.2** for each base station move as *direction* and *distance* required

Fig. 9. The CoMAEC algorithm

In clustering phase, classical data clustering techniques [21] can be used to help base station head to classify the source sensor nodes (the number of cluster equals to the number of base station, and the linkage metric set as the distance between sensor nodes).

## VI. SIMULATIONS

In this section, we present the results of our simulation of the proposed energy conserving method, in comparison with other way of movement of base station. All these algorithms are evaluated with different simulation parameters.

### A. Environment

Simulation environment is the same as the test in section III. The tunable parameters in our simulation are as follows. (1) Number of hotspots $N_h$. Large $N_h$ will degrade the performance of MAEC because it is hard for base station to move close to all hotspots. Therefore, we vary $N_h$'s value from 1 to 7. (2) Number of base station $N_b$. The CoMAEC achieve good performance with large $N_b$. Hence, we use 1 to 4 as its values.

The performance metrics are average, max and min energy consumption. Hotspots are generated randomly locating around base station(s).

### B. Effectiveness of MAEC and CoMAEC

Figure 10 shows the performance of MAEC. We get to know that when there are just a few hotspots MAEC achieves best, and the performance of MAEC converges to that of static base station as the number of hotspots increase (as we have mentioned in section V). And both are much better than that of random move and circle around the boundary.

Figure 11 is the average energy consumption of CoMAEC. We can easily know that CoMAEC make use of base station set much better that the stay still method.

Being few methods on energy control in event driven networks exist, we only contrast the performance of our methods and other way of movement of base station. It is in our belief that when combined with energy-aware routing protocols for event driven networks, our proposal will further reduce the energy consumption (spend on route discovery and maintenance process).

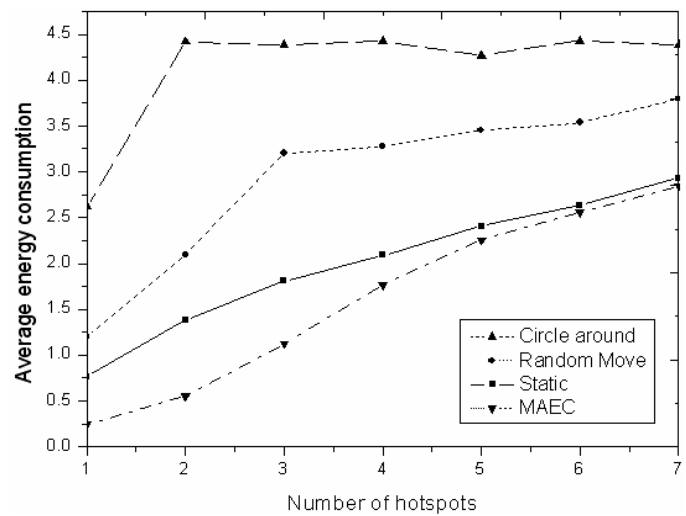

(a) Average energy consumption

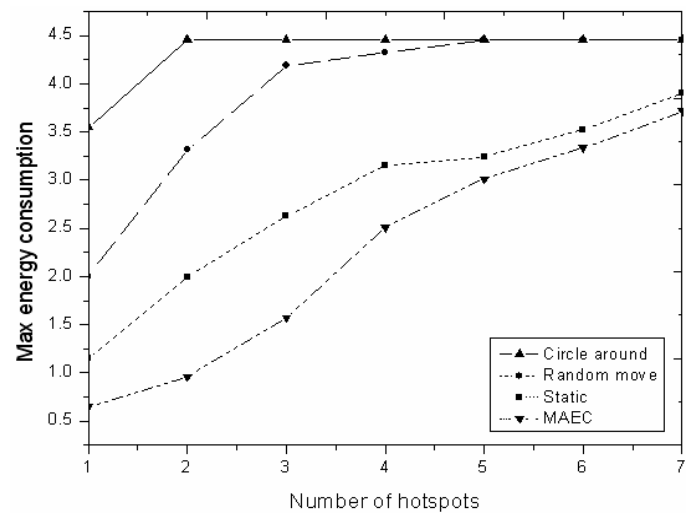

(b) Maximum energy consumption

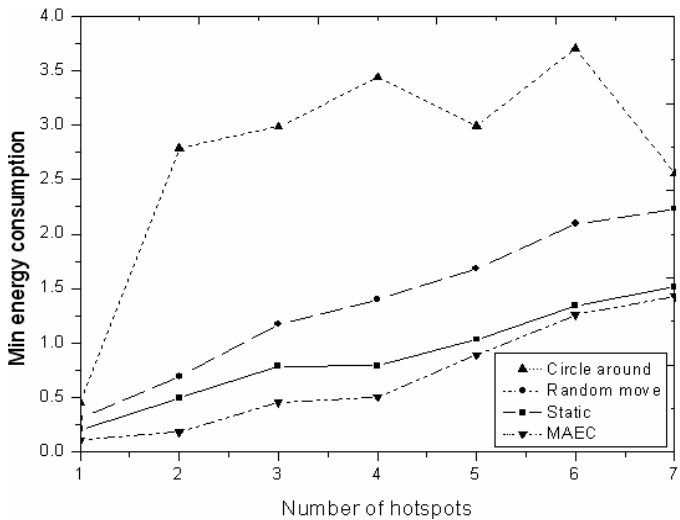

(c) Minimum energy consumption

Fig. 10. Comparison of different performance metric between MAEC and several other movement methods

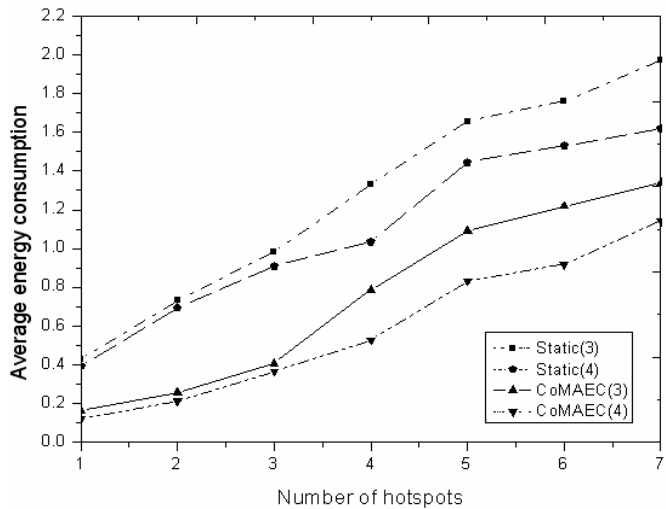

Fig. 11. Comparison of average energy consumption between CoMAEC and static base station

## VII. CONCLUSIONS

To prolong the network lifetime of event driven sensor network applications where data does not need to be gathered continuously, we must make sure that least number of sensor node need to forward data. In this paper, we have first shown that, with base station moving close to the hotspots, the energy consumption of an event driven sensor network falls dramatically. Then we propose a movement-assisted energy conserving method, which direct the way base station moves. We have also considered a cooperative algorithm to make use of base station set to further optimize the data collection. Simulation results show that the proposed method can achieve satisfying effect with modest costs.

We are currently conducting more simulations to validate proposed methods and working over the effect of time interval (*discovery_interval*, *moving_interval*) and velocity on the performance of MAEC.

In terms of future work, we intend to study how routing and data aggregation protocol can be efficiently integrated into the proposed schema.